\documentclass[11pt]{article}
\usepackage{amssymb}
\usepackage{amsmath} 

\newcommand{\bra}{\langle}
\newcommand{\ket}{\rangle}
\newcommand{\weak}{{\tiny \rm{w}}}
\newcommand{\vect}[1]{\mbox{\boldmath $ #1 $}}

\begin{document} 
\baselineskip 6.0mm 
\hfill 
2011 October 8
\\ \hfill
arXiv:1110.xxxx 
\\ \hfill
[subject classes: hep-ph, quant-ph]
\\
\vspace*{12mm}
\begin{center}
{\bf \Large
Apparent Superluminal Muon-neutrino Velocity 
\vspace{2mm} \\
as a Manifestation of Weak Value}
\vspace{6mm}

Shogo Tanimura\footnote{{\tt E-mail: tanimura[AT]is.nagoya-u.ac.jp}}
\vspace{6mm}

{\it
Department of Complex Systems Science,
Graduate School of Information Science
\\
Nagoya University,
Nagoya 464-8601, Japan
}
\vspace{12mm}

Abstract
\vspace{3mm}

\begin{minipage}{110mm}
\baselineskip 5.0mm 
The result of the OPERA experiment revealed that
the velocity of muon-neutrinos was larger than the speed of light.
We argue that this apparent superluminal velocity can be interpreted as a weak value, 
which is a new concept recently studied in the context of quantum physics.
The OPERA experiment setup forms a scheme
that manifests the neutrino velocity as a weak value.
The velocity defined in the scheme of weak measurement can exceed the speed of light.
The weak velocity is not a concept associated to a single phenomenon
but it is a statistical concept 
defined by accumulating data at separated places and by comparing the data.
Neither information nor physical influence is conveyed at the weak velocity.
Thus the superluminal velocity in the sense of weak value
does not contradict the causality law.
We propose also a model for calculating the neutrino velocity
with taking neutrino oscillation into account.
\end{minipage}
\vspace*{20mm}

\begin{minipage}{120mm}
\baselineskip 5.0mm 
Keywords: 
OPERA neutrino experiment,
muon neutrino,
superluminal velocity,
weak value,
weak measurement,
neutrino oscillation,
causality
\end{minipage}

\end{center}

\newpage

\section{Introduction}
Recently the OPERA experiment group \cite{OPERA} reported that
the measured velocity of muon neutrinos was larger than the velocity of light.
Many researchers are proposing 
various interpretations and explanations on this anomalous result.

Here I would like to propose another interpretation on the OPERA experiment.
This will give a new but rather conservative solution of the problem.
I would like to show that
the superluminal velocity of neutrinos is a manifestation of {\it weak values}.
The idea of weak value has been proposed by
Aharonov, Albert, and Vaidman 
many years ago \cite{Aharonov1988, Aharonov1990, Quantum Paradoxes}.
In the scheme of {\it weak measurement},
it can happen that 
the average $ \bra A \ket_\weak $
of measured values of a some physical quantity $ A $
is larger than the largest eigenvalue of $ A $.
If the scheme of the CNGS (CERN neutrino beam to Gran Sasso) experiment 
realizes a weak measurement scheme,
it is possible that the average of the neutrino velocity exceeds the speed of light.
This interpretation can be proved in the ordinary framework of quantum mechanics.
More fortunately, it does not make contradiction with the causality law.

\section{Weak value}
Here I give a brief explanation of the concept of weak value 
with referring to the formulation by Hosoya and  Koga \cite{Hosoya}.
The equation defining the average of an observable $ A $ 
in a given state $ \phi $ is rearranged as
\begin{eqnarray}
	\bra \phi | A | \phi \ket
&=&
	\int d \chi \,
	\bra \phi | \chi \ket \, \bra \chi | A | \phi \ket
	\nonumber \\
&=&
	\int d \chi \,
	\bra \phi | \chi \ket \, \bra \chi | \phi \ket \, 
	\frac{ \bra \chi | A | \phi \ket }{ \bra \chi | \phi \ket }
	\nonumber \\
&=&
	\int d \chi \,
	\Big| \bra \chi | \phi \ket \Big|^2 \, 
	\frac{ \bra \chi | A | \phi \ket }{ \bra \chi | \phi \ket }
	\nonumber \\
&=&
	\int d P ( \chi ) \, A_\weak ( \chi ),
	\label{average}
\end{eqnarray}
where an complete orthonormal set 
$ \int d \chi \, | \chi \ket \bra \chi | = 1 $
is inserted and
$  d P ( \chi ) = | \bra \chi | \phi \ket |^2 \, d \chi $
is the transition probability from the initial state $ \phi $ to a final state $ \chi $.
The number 
\begin{eqnarray}
	A_\weak ( \chi ) :=
	\frac{ \bra \chi | A | \phi \ket }{ \bra \chi | \phi \ket }
	\label{weak value}
\end{eqnarray}
is called a weak value of $ A $
between the two state $ \phi $ and $ \chi $.
In general, the weak value $ A_\weak ( \chi ) $ is a complex number
even if $ A $ is an hermite operator.
Different interpretations have been given to
the real part and the imaginary part of $ A_\weak ( \chi ) $.

Let $ a_{\max} $ and $ a_{\min} $
be the supremum and the infimum of the spectrum of $ A $ respectively,
assuming that $ A $ is a bounded operator.
In general, the weak value $ A_\weak ( \chi ) $ is {\it not} restricted
in the range
$ | A_\weak ( \chi ) | \le | a_{\max} |, | a_{\min} | $.
In particular, when the denominator of (\ref{weak value}) is small,
the weak value can become larger 
than the eigenvalues of $ A $ as shown by Aharonov \cite{Aharonov1988}.
In other words, for events of small probability, we can observe a large weak value.
Of course, by summing up all the possible outcomes, 
we get the ordinary average
$ \bra \phi | A | \phi \ket $,
which is a real number and is in the range
$ a_{\min} \le \bra \phi | A | \phi \ket  \le a_{\max} $.
Although the definition of the weak value may look strange or artificial,
weak values have been measured 
in various experiments of quantum optics \cite{Solli, Yokota, Tamate}.

\section{Implication to the OPERA experiment}
It should be emphasized that
the definition of the weak value refers to 
both the initial state $ \phi $ and the final state $ \chi $,
which are different each other in general.
It also should be emphasized that
the weak value is a statistical quantity.
The weak value is not determined in a single-event experiment.
In a real experiment, we cannot fix the final state.
Usually we repeat experiment runs and
measure various outcomes with various final states.
We calculate the weak value from accumulated data by selecting the final state.
We can observe a large weak value only through rare events,
which have small occurrence probability.
For this feature, a scheme for measuring weak values is called
weak measurement.

These characteristics of the general weak measurement fit
the CNGS neutrino velocity measurement scheme.
In the CNGS beam,
the initial state is defined in terms of 
the proton beam, or the pion/kaon beam at CERN.
The final state is defined in terms of the neutrino scattering events
at the OPERA detector.
Not all of the neutrinos are detected 
because the interaction of neutrinos is really weak in a usual sense.
They observe only a restricted class of events.
The most probable events are that 
the neutrinos fly through the detector without any scattering.
It also should be noted that
the velocity of an individual neutrino 
was not measured in the OPERA experiment.
The neutrino velocity was defined by a statistical method from accumulated data.
These features of the OPERA experiment 
perfectly conform to the weak measurement scheme.

\section{Model}
Let us discuss a simple model for calculation of the average velocity.
Using the Dirac spinor $ \psi $ in the equation
\begin{eqnarray}
	i \frac{\partial}{\partial t} \psi =
	( \vect{\alpha} \cdot \vect{p} + \beta m ) \psi
	\label{Dirac}
\end{eqnarray}
(we use the unit system in which $ c = \hbar = 1 $),
we can define an expectation value of the velocity as
\begin{eqnarray}
	\bra \vect{v} \ket :=
	\frac{ \psi^\dagger \vect{\alpha} \psi }{ \psi^\dagger \psi }.
	\label{average velocity}
\end{eqnarray}
The alpha matrix $ \alpha_i $ has $ \pm 1 $ as its eigenvalues.
In other words, the eigenvalue of the velocity operator is the speed of light.
For a plane wave solution $ \psi \sim e^{-ikx} $,
we get the ordinary velocity 
\begin{eqnarray}
	\bra \vect{v} \ket =
	\frac{ \vect{p} }{ E },
	\label{average velocity of plane wave}
\end{eqnarray}
which does not exceed the speed of light.

The weak value of the velocity is defined as
\begin{eqnarray}
	\vect{v}_\weak :=
	\frac{ \quad \chi^\dagger \vect{\alpha} \phi \quad }{ \chi^\dagger \phi }
	\label{weak velocity}
\end{eqnarray}
between the initial state spinor $ \phi $ and the final state spinor $ \chi $.
When the absolute value of the inner product $ \chi^\dagger \phi $ is small,
the weak velocity can exceed the speed of light.

To determine the spinors $ \phi, \chi $,
we may use a simple model of neutrino oscillation:
\begin{eqnarray}
	i \frac{\partial}{\partial t} 
	\begin{pmatrix}
	\psi_\mu \\ \psi_\tau
	\end{pmatrix}
	=
	( \vect{\alpha} \cdot \vect{p} \otimes 1 
	+ \beta \otimes M ) 
	\begin{pmatrix}
	\psi_\mu \\ \psi_\tau
	\end{pmatrix},
	\label{Dirac osc}
\end{eqnarray}
where 1 and $ M $ are $ 2 \times 2 $ matrices, 
which act on the flavor indices $ \mu, \tau $.
The initial state $ \phi $ of the neutrino 
is an eigenstate of the electroweak interaction
and propagates obeying the equation (\ref{Dirac osc}).
A scattering event at the detector defines the final state wave function $ \chi $.
Then the weak value of the velocity is calculated with (\ref{weak velocity}).
A detailed analysis is necessary to lead a definite consequence.

\section{Causality is not violated}
Can the neutrino with the superluminal weak velocity 
violate the causality law?
In general,
the weak value is determined by data accumulated
in a series of repeated runs of experiment
with a fixed initial state and a selected final state.
The causality law forbids an individual neutrino to convey physical influence 
faster than the speed of light.
But the weak velocity is a statistical quantity 
that is determined by comparing the data accumulated at separated places.
This does not imply that any information of influence is conveyed
at the weak velocity.
Hence, the OPERA experiment does not contradict the relativistic causality.

It also should be noted that
the velocity of a particle during quantum tunneling effect
can be defined in the sense of weak value
and it is allowed to exceed the speed of light \cite{Sokolovski}.

\section*{Acknowledgements}
The author would like to thank the members of Tempaku seminar,
especially,
K.~Hasebe,
S.~Kitakado,
A.~Nakayama,
Y.~Ohnuki,
T.~Okubo,
K.~Sakabe,
T.~Shimizu,
K.~Takano,
A.~Tokumitsu,
M.~Yamanoi,
who participated in the discussion on the superluminal neutrino problem.
This work is supported 
by Japan Society for the Promotion of Science, 
Grant No.~22540410.

\end{document}